# A Circular Crested Lamb Wave Resonator with Spurious Mode Suppression and Quality Factor Enhancement

Xianzheng Lu[1], Liang Lou[2,3,**], and Hao Ren[1,*]

[1] School of Information Science and Technology, ShanghaiTech University, Shanghai, 201210, China
[2] School of Microelectronics, Shanghai University, Shanghai 201800, China
[3] Shanghai Industrial µTechnology Research Institute, Shanghai 201899, China
[*]Corresponding author E-mail: renhao@shanghaitech.edu.cn
[**]Corresponding author E-mail: liang.lou@sitrigroup.com

**Abstract:**
To date, nearly all the reported Lamb wave resonators (LWRs) are straight crested LWRs, which suffer from inherent spurious modes and low quality factors (*Q*). For the first time, this work demonstrates a circular crested LWR. Its advantages over the straight crested LWR are presented comprehensively by studying their fundamental symmetric (S0) mode, which is the simplest and most representative Lamb wave mode. Utilizing circular crested Lamb waves, the proposed resonator avoids only utilizing waves propagating in the lateral direction in the straight crested LWRs, thus eliminating the transverse spurious modes as no transverse direction exists. Besides, different from straight crested Lamb waves maintaining the same displacement amplitude along the propagation direction, circular crested Lamb waves exhibit displacement attenuation toward device edges, which effectively concentrates energy in the device center, assists in reducing energy loss through anchors, and improves *Q*. Based on 20 at.% scandium-doped aluminum nitride ($Al_{0.8}Sc_{0.2}N$) thin films, the microfabricated circular crested LWR effectively suppresses transverse spurious modes and achieves a 40.7% *Q* improvement in experiments with no degradation in effective electromechanical coupling coefficients ($k_{eff}^2$) compared with the conventional straight crested LWR when working in S0 mode in experiments, in contrast with the conventional straight crested LWR which shows inherent experimental transverse spurious modes. Moreover, the free edges covered by top electrodes enhance device robustness against misalignment and over-etching in the fabrication process. With the advantages of spurious mode suppression, *Q* enhancement, and fabrication robustness, the circular crested LWR is a promising candidate for next-generation filters and oscillators.

**Keywords**: Piezoelectric resonators, Circular crested Lamb wave, Lamb wave resonators, Spurious mode suppression, Quality factor enhancement.

## 1. Introduction

Wireless communication technologies have shown rapid development in the past few decades. From 1G to 5G, the fast-evolving speed and quality of communication as well as the trend of miniaturized communication devices come with demanding requirements for high-speed, small-footprint, and low-loss radio frequency (RF) devices to perform timing and filtering [1]. In recent years, Lamb wave resonators

(LWRs) have become a promising candidate for constructing next-generation RF devices due to their small footprints, various excitable modes with a large range of phase velocities, and lithography-defined operation frequencies, which enable megahertz to tens of gigahertz applications and multi-frequency on one chip integrations [2-11].

To date, nearly all the previously reported LWRs are straight crested LWRs based on rectangular interdigitated transducers (IDTs) utilizing laterally propagating Lamb waves, yet they suffer from various inherent spurious modes and low quality factors ($Q$) [12-15]. These two disadvantages severely prevent straight crested LWRs from constructing high-performance RF devices, as spurious modes can result in unwanted pass or stop bands for filters, and a low $Q$ means high phase noise for oscillators as well as high insertion loss and low out-band rejection for filters [16, 17]. As a result, how to achieve spurious mode suppression and $Q$ enhancement for straight crested LWRs has been widely studied by researchers. Notably, straight crested LWRs based on the fundamental symmetric (S0) mode are the most focused on because they exhibit the simplest and most representative mode shape, barely dispersive phase velocity, and excitation effectiveness, which render them an excellent example for the study on spurious mode suppression and $Q$ enhancement [12, 13, 18-23].

Although various excitable modes bring straight crested LWRs broad application scopes, they can also cause challenging spurious mode issues when a certain mode is needed [24, 25]. For S0 mode straight crested LWRs, transverse spurious modes are a representative category. They are mainly due to the additional unwanted wave components with different wavenumbers along the transverse direction, which result from the insufficient wave guiding of IDTs caused by wave velocity mismatch of the IDT, gap, and busbar regions in the transverse direction [26]. Besides, tethers and free edges beside tethers in the transverse direction also influence the behavior of transverse wave components and thus the excitation of transverse spurious modes [27]. As a result, as long as straight crested LWRs with rectangular IDTs and resonance cavities are utilized, transverse spurious modes are inherent, and additional design modifications for transverse spurious mode suppression are required. Many researchers have proposed different methods for transverse mode suppression for straight crested LWRs, mostly focusing on modifying boundary conditions in the transverse direction. Yantchev *et al.* effectively suppressed transverse spurious modes of S0 mode AlN straight crested LWRs by adding additional matching regions between the IDT and busbar regions [26]. However, this method needs precise control of the matching region geometries to ensure the suppression effectiveness. Zhang *et al.* reported that extending the gap length between the IDT and busbar region could also mitigate transverse spurious mode, yet this method is at risk of introducing new spurious modes [27]. Utilizing vertical electrode protrusions or lateral resonant cavity protrusions at the IDT finger ends was also demonstrated to be an effective suppression method by Zhang *et al.* [28]. Nevertheless, this method requires a more complicated fabrication process, which can increase the fabrication cost and probability of defects. Zou *et al.* proposed adding hammer head-shape regions at IDT finger ends to realize the piston mode for transverse spurious mode suppression. However, it also requires accurate designs to effectively suppress transverse spurious modes, otherwise additional spurious modes can be excited [29, 30]. Zhu *et al.* reported that applying the top

IDT apodization technique was another effective approach, yet it is at the cost of degrading $k_{eff}^2$ [24].

To enhance $Q$ of straight crested LWRs, reducing energy loss is an extensively employed strategy. Taking S0 mode straight crested LWRs as an example, various $Q$ enhancement methods through diminishing energy dissipation to the surroundings have been proposed. Lin *et al.* [31] and Wu *et al.* [32] utilized phononic crystal (PnC) designs on tethers of S0 mode straight crested LWRs to reduce energy loss through tethers for higher $Q$ and successfully obtained $Q$ improvements of 38% at 494.43 MHz and 45.2% at 1.0198 GHz respectively. However, PnCs with a certain geometry can only realize energy loss reduction at specific frequency bands. Besides, small fabrication deviations may lead forbidden bands of PnC to drift away from the resonant frequency of LWRs. Biconvex free edges in the wave propagation direction were utilized by Lin *et al.* [33] and Siddqi *et al.* [34] to replace flat edges for $Q$ enhancement on AlN and AlN-on-Si platforms respectively. Although Lin *et al.* successfully achieved 161.3% $Q$ improvement and Siddiqi *et al.* achieved nearly 600% unloaded quality factor ($Q_u$) enhancement, their straight crested LWRs both suffered $k_{eff}^2$ degradation of approximately 50%. Butterfly-shaped edges beside the tethers are another method for $Q$ improvement, by which around 50% enhancement in $Q$ was achieved in S0 mode AlN and 30 at. % scandium doped AlN ($Al_{0.7}Sc_{0.3}N$) straight crested LWRs by Zou *et al.* [35, 36] and Luo *et al.* [23] respectively without negatively affecting $k_{eff}^2$. Nevertheless, how to determine the most effective butterfly-shaped edge design remains to be further explored.

Although the aforementioned approaches effectively suppress spurious modes or enhance $Q$, few of them can simultaneously achieve both of the optimization goals. Furthermore, they complicate the design procedure of straight crested LWRs and may interfere with the optimization of other performances other than spurious mode suppression and $Q$. As a result, designing a new LWR topology without utilizing straight crested Lamb waves is urgently needed. In this work, we report a circular crested LWR for the first time. To demonstrate its advantages over conventional straight crested LWRs, the S0 mode of the two types of LWRs is investigated and compared, as it is the most representative mode. Different from the straight crested Lamb waves utilized by conventional LWRs, circular crested Lamb waves propagate radially and the transverse direction does not exist, thus eliminating transverse spurious modes. Besides, the circular crested Lamb waves experience fast displacement magnitude reduction in the first several wavelengths [37], thus concentrating energy in the device center. Moreover, we design zero spaces between the outermost IDT fingers and free edges, which provide resistance against fabrication deviations such as misalignment and over-etching. Compared with the conventional straight crested LWR, the fabricated circular crested LWR successfully achieves the suppression of transverse spurious modes near the main S0 mode resonance as well as a 40.7% $Q$ improvement with no $k_{eff}^2$ degradation.

## 2. Operation Principles and Theoretical Modeling
### 2.1 Wave Motion of S0 Mode Lamb Waves on Thin Plates

To illustrate the advantages of utilizing S0 mode circular crested Lamb waves, we start with the wave equation of Lamb waves. As shown in Figure 1, in a lossless infinite thin plate whose top and bottom surfaces are traction-free, the displacements of a symmetric mode straight crested Lamb wave propagating in the $x$ direction are given by [37]:

$$\begin{cases} u_x = i\big(\xi A\cos(\alpha z) + \beta B\cos(\beta z)\big)e^{i(\xi x-\omega t)} \\ u_z = -\big(\alpha A\sin(\alpha z) - \xi B\sin(\beta z)\big)e^{i(\xi x-\omega t)} \end{cases} \qquad (1)$$

where

$$\begin{cases} \alpha^2 = \dfrac{\omega^2}{c_p^2} - \xi^2 \\ \beta^2 = \dfrac{\omega^2}{c_s^2} - \xi^2 \end{cases}, \qquad (2)$$

$$\frac{A}{B} = \frac{(\xi^2 - \beta^2)\sin(\beta d)}{2\xi\alpha \sin(\alpha d)}, \qquad (3)$$

where $u_x$ and $u_z$ are displacements in $x$ and $z$ directions respectively, $\omega$ is the wave angular frequency, $\xi$ is the wavenumber in the wave propagation direction, $\alpha$ and $\beta$ are the $z$-direction wavenumbers of dilatational and shear waves, $c_p$ and $c_s$ are the phase velocities of dilatational and shear waves, and $d$ is half of the plate thickness. As for symmetric mode circular crested waves, the displacements are given by the cylindrical coordinate [38]:

$$\begin{cases} u_r = \left(\dfrac{\xi}{\alpha^2 - \beta^2}C\cosh(\alpha z) - \dfrac{\beta}{\xi}D\cosh(\beta z)\right)J_1(\xi r)e^{i\omega t} \\ u_z = \left(-\dfrac{\alpha}{\alpha^2 - \beta^2}C\sinh(\alpha z) + D\sinh(\beta z)\right)J_0(\xi r)e^{i\omega t} \end{cases} \qquad (4)$$

where $u_r$ is the displacement along the radius direction, $J_n(x)$ is the $n$th Bessel function with

$$J_n(x) = \sum_{m=0}^{\infty} \frac{(-1)^m \left(\frac{x}{2}\right)^{2m+n}}{m!(m+n)!}, \qquad (5)$$

and

$$\frac{C}{D} = \frac{(\alpha^2 - \beta^2)(\xi^2 + \beta^2)\sinh(\beta d)}{2\alpha\xi^2 \sinh(\alpha d)}. \qquad (6)$$

For S0 mode Lamb waves propagating in thin plates with thickness much smaller than their wavelengths, they exhibit the low-frequency long-wavelength behavior, which is described by $z$-direction wavelengths $2\pi/\alpha$ and $2\pi/\beta$ far beyond half the plate thickness $d$ [37]. In this way, $\alpha z$ and $\beta z$ with $-d \leq z \leq d$ are approximately zero, which reduces the displacements of straight crested Lamb waves to

$$\begin{cases} u_x = iB\left(\dfrac{\xi^2 - \beta^2}{2\alpha} + \beta\right)e^{i(\xi x - \omega t)} \\ u_z = 0 \end{cases} \tag{7}$$

and the displacements of circular crested Lamb waves to

$$\begin{cases} u_r = D\left(\dfrac{\xi^2 + \beta^2}{2\alpha\xi} - \dfrac{\beta}{\xi}\right)J_1(\xi r)e^{i\omega t} \\ u_z = 0 \end{cases} \tag{8}$$

In this way, the $z$-direction displacements are approximately zero while the $x$- and $r$-direction displacements are approximately invariant with $z$, as shown in Figure 2. It can be observed that the displacement amplitude of a circular crested Lamb wave rapidly reduces by more than 60% after propagating two wavelengths, while that of a straight crested Lamb wave remains invariant with the distance from the origin. In other words, for a circular crested and a straight crested Lamb wave carrying the same amount of energy, the former concentrates more energy in the geometric center of the wave train while the latter evenly distributes the energy along the whole wave train.

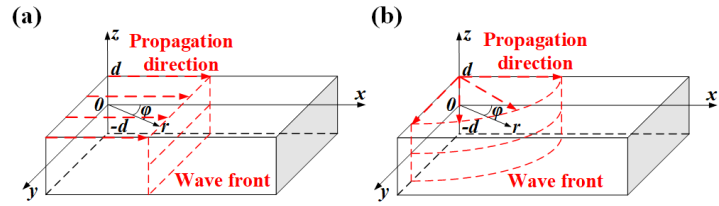

**Fig. 1.** A schematic of the infinite thin plate with thickness $2d$ where (a) straight crested and (b) circular crested Lamb waves propagate.

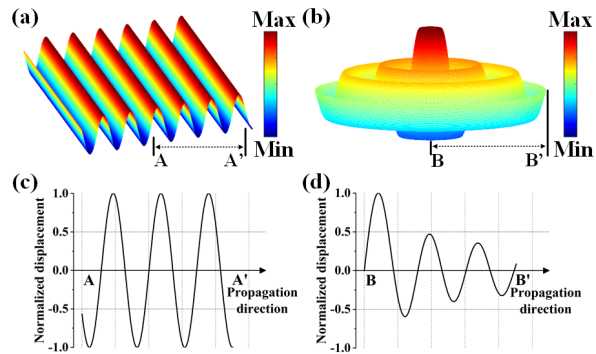

**Fig. 2.** 3D surface diagrams of (a) the normalized $x$-direction displacement of a straight crested Lamb wave and (b) the normalized $r$-direction displacement of a circular crested Lamb wave on a lossless infinite plate with half of the plate thickness $d$ far smaller than $2\pi/\alpha$ and $2\pi/\beta$; (c) the displacement magnitude of the straight crested Lamb wave along $AA'$; (d) the displacement magnitude of the circular crested Lamb wave along $BB'$.

## 2.2 Device design of S0 mode LWRs

In piezoelectric thin films, Lamb waves can be excited through the inverse piezoelectric effect. For example, the inverse piezoelectric effect in AlN and AlScN can be expressed by

$$\begin{bmatrix} S_{xx} \\ S_{yy} \\ S_{zz} \\ S_{yz} \\ S_{zx} \\ S_{xy} \end{bmatrix} = \begin{bmatrix} 0 & 0 & d_{31} \\ 0 & 0 & d_{31} \\ 0 & 0 & d_{33} \\ 0 & d_{15} & 0 \\ d_{15} & 0 & 0 \\ 0 & 0 & 0 \end{bmatrix} \begin{bmatrix} E_x \\ E_y \\ E_z \end{bmatrix}, \qquad (9)$$

where $S$ is the strain, $d_{ij}$ is the inverse piezoelectric coefficient, and $E$ is the electric field. As illustrated in (7) and (8), the S0 mode Lamb wave displacements are mainly in-plane displacements when the thickness of the thin plate carrying waves is far smaller than the wavelength. Thus, for piezoelectric thin plates fabricated with materials that can only utilize vertical electric fields to induce in-plane strains, such as AlN and AlScN, properly controlling vertical electric fields is the key to generating desired S0 mode Lamb waves. Arranging the configuration of electrodes is an effective approach, by which the locations with maximum vertical electric fields can be defined. The in-plane strain at these locations is also maximum according to (9), and thus the distribution of S0 mode Lamb waves can be determined. Top IDTs are widely employed to define the electric field distributions due to the convenience of implementing electrical connection and patterning. Besides, applying additional bottom floating electrodes can effectively constrain electrical fields in the vertical direction and enhance coupling efficiency [39]. As an example, structures consisting of top IDTs, $Al_{0.8}Sc_{0.2}N$ layers, and bottom floating electrodes are chosen for the proposed S0 mode straight crested and circular crested LWRs, as shown in Figure 3. In the conventional straight crested LWR, straight IDT fingers are arranged periodically in the lateral direction for exciting straight crested S0 mode Lamb waves. As for the circular crested LWR, circular IDT fingers are located periodically in the radial direction to excite S0 mode circular crested Lamb waves.

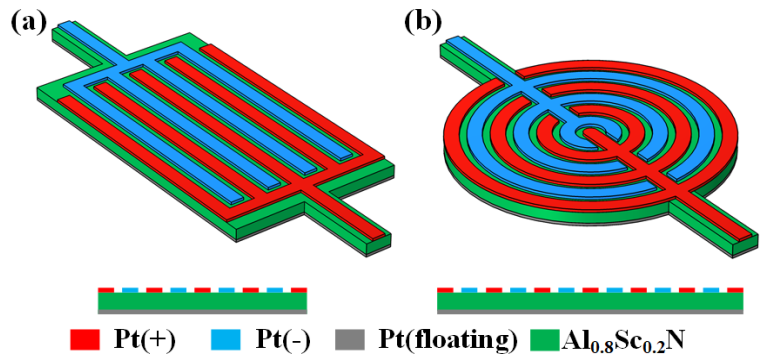

**Fig. 3.** Schematics of (a) the proposed S0 mode straight crested and (b) circular crested LWRs.

TABLE 1

Design Parameters of Ideal-Design LWRs in Simulation

| Parameters | Straight Crested | Circular Crested |
|---|---|---|
| $W_{elec}$ | 10 μm | 10 μm |
| $W_{space}$ | 5 μm | 5 μm |
| $W_{tether}$ | 1 μm | 1 μm |
| $t_{top}$ | 0.2 μm | 0.2 μm |
| $t_{piezo}$ | 1.1 μm | 1.1 μm |
| $t_{btm}$ | 0.2 μm | 0.2 μm |
| $r_{su}$ | 60 μm | 60 μm |
| $r_{PML}$ | 30 μm | 30 μm |

## 3. FEA Simulation of LWRs
### 3.1 LWRs with ideal designs

In LWRs, Lamb waves exist in lossy finite thin plates instead of lossless infinite thin plates, in which the additional traction-free and fixed boundary conditions in the propagation plane can influence the wave characteristics, such as wave attenuation and permitted wave modes [40]. Moreover, it is challenging to obtain analytical solutions for general Lamb waves on finite thin plates [38]. Thus, finite element analysis (FEA) by COMSOL is utilized to analyze the wave motion on lossy finite thin plates of the proposed circular crested and straight crested LWRs. Resonators with ideal designs are first simulated, whose design parameters are illustrated in Table 1. As shown in Figure 4(a) and (b), the busbars connecting the IDT fingers are omitted to minimize the influence of electrode nonideality on the excitation of S0 mode Lamb waves. Besides, the tether widths are set to 1 μm to minimize the distortion of plate shapes and interference of Lamb waves. The suspended surroundings caused by over-etching in the release step of the fabrication process are also taken into consideration to more accurately simulate the fixed boundary condition of the resonator [41]. The simulated $u_x$ of the straight crested LWR and $u_r$ of the circular crested LWR are shown in Figure 4(c) and (d) respectively, and the displacement variations along the *x*-axis at the device centers are shown in

Figure 4(e). There is no transverse direction for the circular crested Lamb wave on the finite thin plate, which may avoid the excitation of transverse spurious modes. In contrast, there is a transverse direction for the straight crested Lamb wave, which can support unwanted transverse spurious modes. Moreover, similar to the infinite plate condition, the circular crested Lamb wave still achieves the maximum displacement magnitude at the plate center, while the displacement magnitude of the straight crested Lamb wave is approximately unchanged along the *x*-direction. In other words, in finite thin piezoelectric plates, the excited circular crested Lamb wave also concentrates more energy in its geometric center, which is also the plate center. As less energy is distributed at device edges, energy leakage to the surroundings can possibly be reduced, which can theoretically result in higher *Q*. In this way, it is theoretically possible to utilize the characteristics of circular crested Lamb waves to benefit LWRs.

To further investigate the effectiveness of utilizing circular crested Lamb waves in LWRs, frequency responses and modes of LWRs with ideal designs shown in Figure 5(a) and (b) are simulated. The simulated admittance versus frequency response curves of the two resonators are shown in Figure 5(a) and (b) respectively. The highest peaks in the admittance curves correspond to the S0 modes of the two resonators, whose mode shapes are shown in Figure 5(c) and (d) respectively. In the spectrum of the ideal-design straight crested LWR, three distinct spurious modes can be spotted near the main S0 mode. As shown in Figure 5(e), the unwanted modes are transverse spurious modes inherent to straight crested LWR, which show additional space periodicity along the transverse direction and are caused by the acoustic impedance mismatch at edges in the transverse direction [26, 27, 30]. On the other hand, the ideal-design circular crested LWR shows no transverse spurious modes and a clean spectrum near the main S0 mode resonance. The reason is that the space periodicity of circular crested Lamb waves distributes in the radius direction, which means that all the directions from the center to the edge of the round resonance cavity are the same. As there is no transverse direction relative to the propagation direction, the transverse spurious modes thus vanish. Besides, two important parameters, $k_{eff}^2$ and fitted quality factor ($Q_{fit}$), which are defined by

$$k_{eff}^2 = \frac{\pi f_s}{2 f_p} \cdot \frac{1}{\tan\left(\frac{\pi f_s}{2 f_p}\right)} \tag{10}$$

and

$$Q_{fit} = \frac{2\pi f_s L_m}{R_m + R_s}, \tag{11}$$

are extracted by fitting the admittance curves to the modified Butterworth-Von Dyke (mBVD) equivalent circuit model. In the model, $C_0$ is the static capacitance, $C_m$ is the motional capacitance, $L_m$ is the motional inductance, $R_m$ is the motional resistance, $R_s$ is the series resistance, $R_0$ is the resistance representing the dielectric loss, and $f_s$ and $f_p$ are the resonant and anti-resonant frequencies. Compared to the straight crested LWR, the ideal-design circular crested LWR achieves a 35.8% enhancement in $k_{eff}^2$, which may be attributed to the suppression of transverse spurious modes, as less energy is coupled to the unwanted spurious modes. Moreover, the circular crested LWR demonstrates a $Q_{fit}$ of 1171.4, which is 15.2% higher than that of the straight crested LWR. The $Q$ enhancement demonstrates the effectiveness of the circular crested LWR in reducing energy dissipation through tethers to suspended surroundings, which is the dominant energy loss at hundreds of MHz [42].

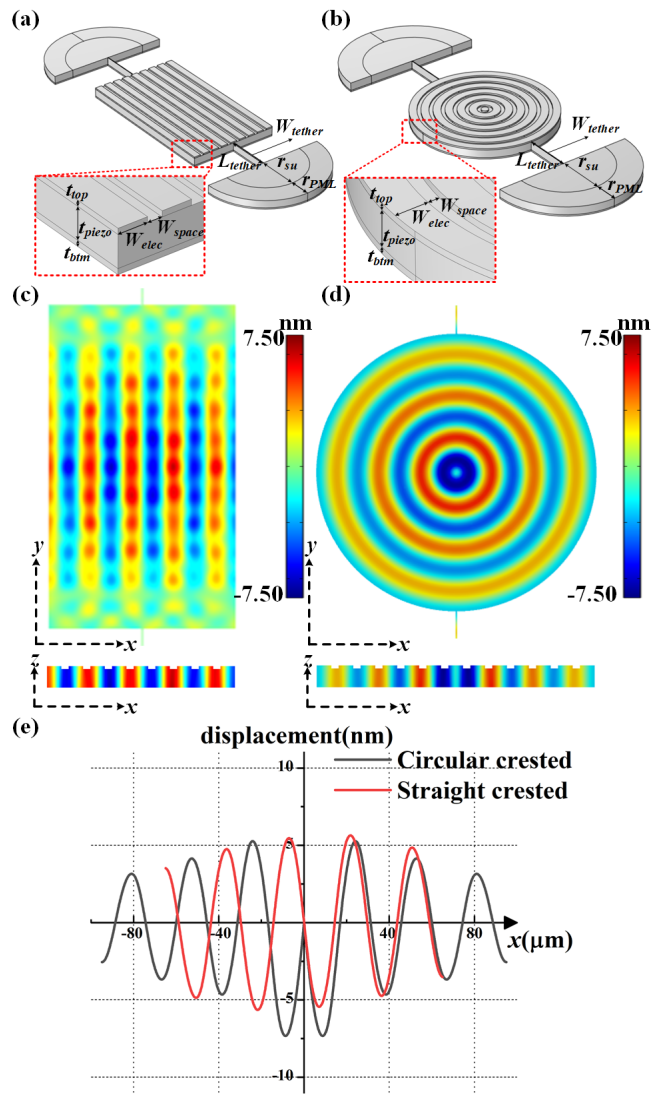

**Fig. 4.** 3D models of the (a) straight crested and (b) circular crested LWRs with ideal designs in COMSOL; the simulated 2D in-plane displacements of the (c) straight and (d) circular crested LWRs with ideal designs; (e) the relationship between the in-plane displacement and $x$ positions at device centers.

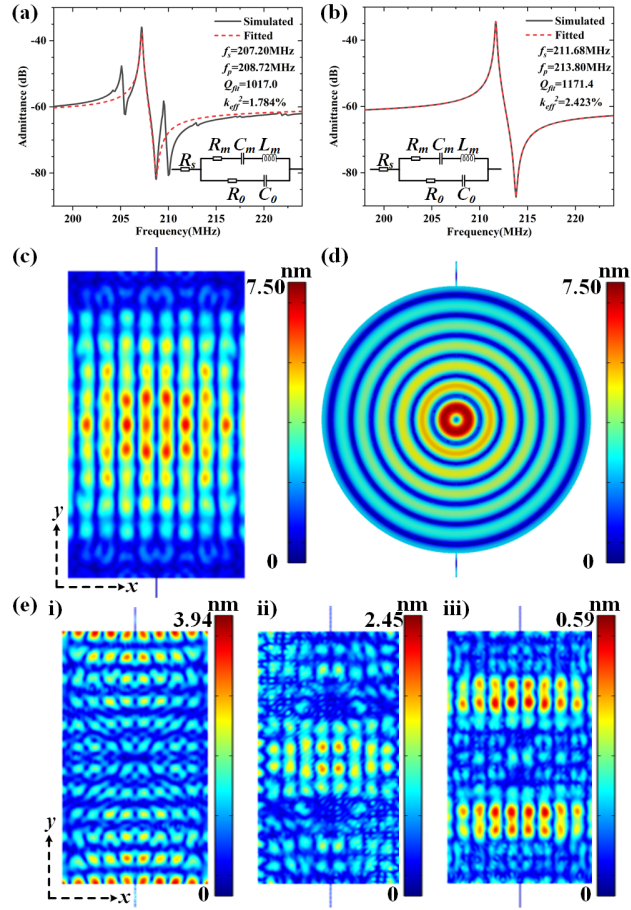

**Fig. 5.** The simulated and mBVD fitted admittance of (a) straight crested and (b) circular crested LWRs with ideal designs; the simulated S0 mode displacement magnitudes of (c) straight crested and (d) circular crested LWRs with ideal designs; (e) the simulated spurious mode displacement magnitudes of the straight crested LWR with ideal design at i) 205.1 MHz, ii) 209.5 MHz, and iii) 212.9 MHz.

### 3.2 LWRs with actual designs

Afterward, the circular and straight crested LWRs with actual designs are simulated. As shown in Figure 6(a) and (b), the actual designs employ wider tethers as they are needed in real devices to maintain structure stability. Besides, busbars required for electrical connection are also considered. Moreover, unwanted rounded corners due to the nonideal fabrication process are included in the 3D model. Other damping sources arising from experimental imperfections but difficult to include in the 3D model, such as out-of-plane device curvatures and material dampings, are simplified by including them in the damping coefficients. The displacements are first simulated to investigate the influence of the nonideal electrodes and tethers on wave motions. As shown in Figure 6(c-e), compared to those of the ideal-design LWRs, the displacement amplitudes of the actual-design straight crested and circular crested LWRs are significantly reduced due to the increased energy leakage to the substrate caused by the tethers with actual width and additional simplified experimental damping sources included in the damping coefficients. Besides, the circular crested LWR shows suppressed displacements along the *y*-axis, as the busbar in the *y*-direction cannot assist in exciting circular crested Lamb waves. On the other hand, in regions other than the *y*-axis, similar to the ideal-design circular crested LWR, the actual-design circular crested LWR shows displacement attenuating with increasing distance from the device center. Besides, compared with the actual-design straight crested LWR, it shows larger displacement magnitudes at the device center, which is similar to the simulation results of ideal-design LWRs.

After confirming the excitement of circular crested Lamb waves on circular crested LWRs with actual designs, 3D simulations are performed to optimize the tether width, as it can strongly affect the performance of LWRs [43]. As shown in Table 2, a tether width of 20 μm is an optimized choice for both achieving a relatively high $Q$ and an acceptable $k_{eff}^2$. The simulated and mBVD fitted admittances of the actual-design straight crested and circular crested LWRs with tethers of 20 μm width are shown in Figure 7(a) and (b), respectively. Notably, the simulated $k_{eff}^2$ of the actual-design straight crested LWR is similar to that of the ideal-design one, which is possibly due to the suppressed transverse spurious modes. As wider tethers are applied, transverse spurious modes suffer increased energy dissipation to the substrate and thus enhanced damping, which can suppress the spurious modes [44]. Besides, the busbar region in the actual design can mitigate the acoustic impedance mismatch between the IDT region and free boundaries, which is similar to the additional matching regions mitigating the mismatch between the IDT and busbar regions in previous studies to suppress transverse spurious modes [26, 45]. The mitigation of transverse spurious modes enables more energy to be coupled to the main S0 mode and maintains the $k_{eff}^2$. On the contrary, the actual-design circular crested LWR shows a $k_{eff}^2$ degraded by 29.6% compared with the ideal-design one, which may be caused by the reduced coupling efficiency resulting from the nonideal structure. With almost equal electrode areas, comparable electric energy is applied to the actual- and ideal-design circular crested LWRs. However, in the fan-shaped regions adjacent to the busbars of the actual-design one, displacements are barely induced whereas busbars apply electric fields. In other words, electrical energy delivered by busbars is hardly converted to mechanical energy, which reduces the coupling efficiency of the main S0 mode and thus results in a degraded $k_{eff}^2$. As a result, the actual-design straight and circular crested LWRs demonstrate approximately equal simulated $k_{eff}^2$. From another perspective, the actual-design circular crested LWR shows larger displacement than the actual-design straight crested LWR in regions away from the busbars, which demonstrates higher effective coupling efficiency in these regions [5]. If the busbars can be optimized to allow enhanced displacement excitation and thus achieve higher effective coupling efficiency in the fan-shaped regions, the actual-design circular crested LWR theoretically can achieve enhanced effective coupling efficiency and thus a higher $k_{eff}^2$ compared with the straight crested one.

Consistent with the simulation results of the ideal-design LWRs, the actual-design circular crested LWR

maintains a clean spectrum around the main S0 mode resonance, while the straight crested LWR with the actual design still suffers from inherent transverse spurious modes near the main S0 mode, whose mode shapes are shown in Figure 7(c). Moreover, the simulated $Q_{fit}$ of the circular crested LWR is 40.6% higher than that of the straight crested LWR, which demonstrates an improved enhancement compared with the simulation results of ideal-design LWRs as the influence of anchor loss is increased with wider tethers. From the S0 mode shapes shown in Figure 7(d) and (e) and the displacement magnitudes of the straight crested LWR along *AA'* and the circular crested LWR along *CC'* shown in Figure 7(f), it can be observed that the circular crested LWR maintains larger displacement magnitudes at the device center while the two LWRs has approximately equal displacements at tethers and suspended surroundings. Larger displacement magnitudes of the circular crested LWR indicate higher stored acoustic energy in the circular crested LWR per cycle of vibration, while comparable displacements at tethers and suspended surroundings of the two LWRs suggest nearly equal lost acoustic energy in the two LWRs per cycle of vibration. As *Q* is positively correlated with the ratio of acoustic energy stored and acoustic energy lost per cycle of vibration [42, 46, 47], the circular crested LWR can obtain higher $Q_{fit}$.

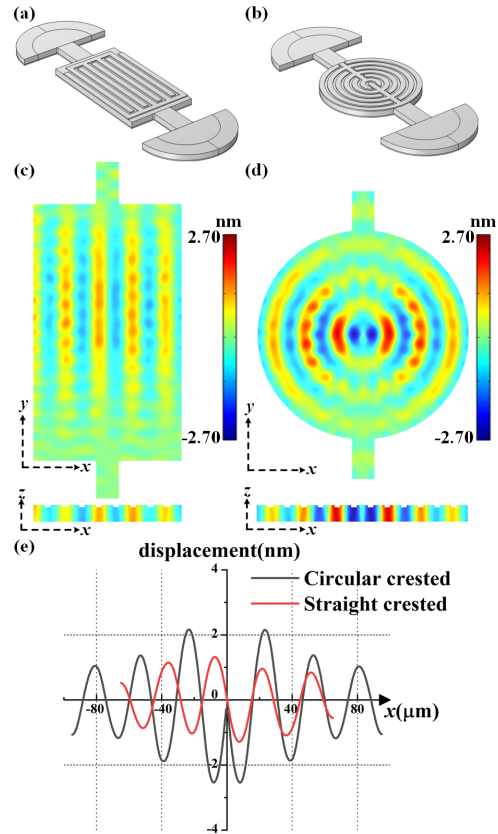

**Fig. 6.** 3D models of the (a) straight crested and (b) circular crested LWRs with actual designs in COMSOL; the simulated 2D in-plane displacements of the (c) straight and (d) circular crested LWRs with actual designs; (e) the relationship between the in-plane displacement and *x* positions at device centers.

TABLE 2

Relationship Between $w_{tether}$ of the Circular Crested LWR and $Q$ and $k_{eff}^2$

| Tether width | $Q$ | $k_{eff}^2$ |
| --- | --- | --- |
| 15 μm | 307.5 | 1.751% |
| 20 μm | 323.0 | 1.705% |
| 25 μm | 301.2 | 1.796% |
| 30 μm | 299.7 | 1.842% |
| 35 μm | 307.4 | 1.877% |

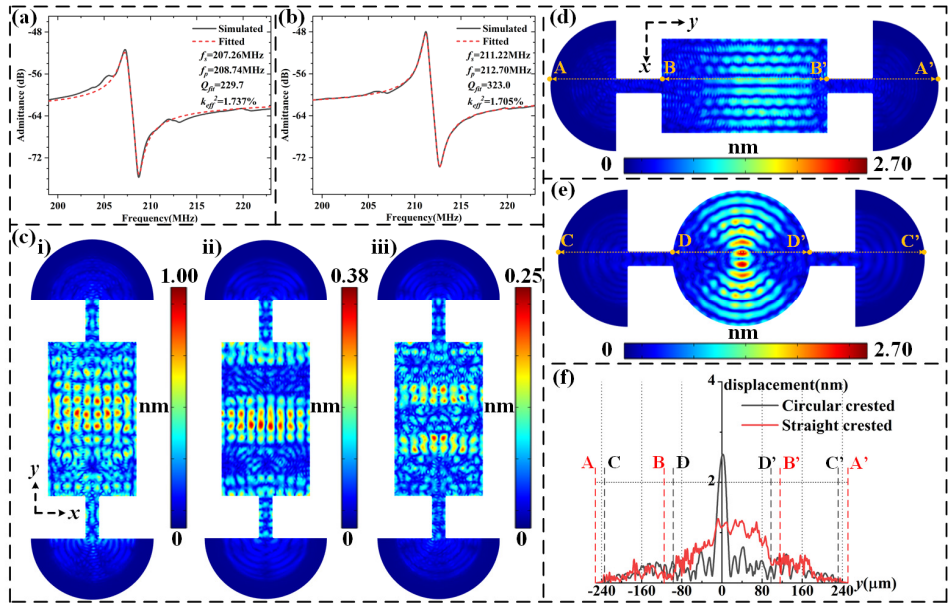

**Fig. 7.** The simulated and mBVD fitted admittances of the (a) straight crested and (b) circular crested LWRs with actual designs; (c) the simulated spurious mode displacement magnitudes of the straight crested LWR at i) 205.5 MHz, ii) 212.0 MHz, and iii) 219.8 MHz, respectively; the simulated S0 displacement magnitudes of the (d) straight crested and (e) circular crested LWRs with actual designs; (f) the relationship between the displacement magnitudes and $y$ positions.

## 4. Device Fabrication and Topography Characterization

To demonstrate the effectiveness of the circular crested LWR design, circular crested and straight crested LWRs are fabricated with 20 at.% scandium-doped aluminum nitride ($Al_{0.8}Sc_{0.2}N$) thin films as the piezoelectric layer. Figure 8 shows the fabrication process, which starts with a 1.1 μm thick $Al_{0.8}Sc_{0.2}N$ thin layer on a 710 μm thick Si wafer. First, the lift-off process is chosen to fabricate the top 200 nm thick Pt electrodes. The image reversal process by AZ5214 photoresist is used to define the electrode region with a maskless aligner (Heidelberg MLA150) and mask aligner (SUSS MA6), followed by the deposition of a 10 nm thick Ti adhesion layer and a 200 nm thick Pt layer through e-beam evaporation (DENTON Explorer-14). After the exfoliation of unwanted metals through N-methyl-2-pyrrolidone (NMP) baths, the top electrodes are formed. Then, a $SiO_2$ hard mask is deposited by plasma enhanced chemical vapor deposition (PECVD) (Oxford PlasmaPro 80) and patterned by reactive ion etching (RIE) (Advance Vacuum Vission 322) with SPR220-3 soft mask. After that, the $Al_{0.8}Sc_{0.2}N$ film is patterned by inductively coupled plasma reactive ion etching (ICP-RIE) (LEUVEN Haasrode-E200A). Remarkably, due to the overlap of the lateral outer edges of the outermost IDT fingers and the resonance cavity in the resonator design, the tolerance of misalignment in the RIE and ICP-RIE step is increased. To assist the backside release process, the wafer is thinned to a thickness of 350 μm from the backside. Afterward, with a backside SPR220-3 soft mask, the Si wafer is etched through from the backside by $XeF_2$ etching (Memsstar Orbis Alpha) to release the resonators. After stripping the backside soft mask, deposition of 10 nm thick Ti and a 200 nm thick Pt from the backside is performed by e-beam evaporation to define the bottom electrodes. Finally, RIE is utilized to remove the $SiO_2$ hard mask on top.

After fabrication, the characteristics of the $Al_{0.8}Sc_{0.2}N$ thin film need to be measured. To determine the actual composition of the $Al_{0.8}Sc_{0.2}N$ thin film, energy dispersive X-ray spectroscopy (EDS) is performed by a scanning electron microscope (SEM) (ZEISS Gemini300). The results in Figure 9(a) show that the ratio between concentrations of aluminum, scandium, and nitride is 4:1:5, which verifies that the piezoelectric thin film is 20 at.% Scandium-doped. Besides, the $Al_{0.8}Sc_{0.2}N$ film surface is characterized by an atomic force microscope (AFM) (Oxford Cypher S) and SEM. As shown in Figure 9(b) and (c), the $Al_{0.8}Sc_{0.2}N$ film presents a low surface roughness deviation ($R_a$) of 3.1 nm while no abnormal orientation grains (AOGs) exist.

The morphology of the fabricated circular crested and straight crested LWRs is characterized by SEM, as shown in Figure 10. The fabricated IDT finger widths are measured to be 11.20 μm and 10.63 μm respectively, demonstrating small deviations from the designed width of 10 μm. On the other hand, undesirable etching can be spotted at LWR edges. To further characterize the nonideal fabrication results and the resonator structures, cross-sectional images are captured by a helium ion microscope (HIM) (ZEISS ORION NanoFab) after cutting an LWR by gallium focused ion beam (Ga-FIB), as shown in Figure 11. Notably, a Pt protection layer is deposited to protect the LWR geometry during the Ga-FIB milling process.

As shown in Figure 11(a), the thicknesses of the top Pt electrode, $Al_{0.8}Sc_{0.2}N$ layer, and bottom Pt electrode are 205 nm, 1.155 μm, and 216 nm respectively, which are consistent with the design parameters. Figure 11(b) shows oblique sidewalls with an angle of 57.8° instead of vertical ones, which may result from the imperfection in the $Al_{0.8}Sc_{0.2}N$ ICP-RIE step and can negatively affect the performance of the fabricated LWRs [25, 48]. Besides, the edge of the outermost IDT finger is 77 nm thick, which results from the misalignment of $SiO_2$ hard mask patterning. Due to the misalignment, the outermost IDT finger edge is first etched by the RIE over-etching of $SiO_2$ hard mask, and then etched by the ICP-RIE step as it is not covered by the hard mask. On the other hand, it indicates that the outermost IDT fingers successfully protect the underneath $Al_{0.8}Sc_{0.2}N$ layer during the RIE and ICP-RIE step, which verifies the robustness against misalignment and over-etching brought by the design of overlapping the lateral outer edges of the resonance cavity and the outermost IDT fingers.

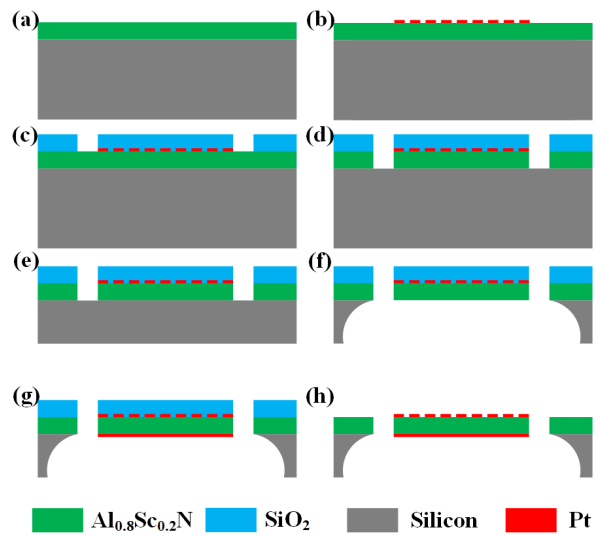

**Fig. 8.** The fabrication process of the circular crested and straight crested LWRs. (a) $Al_{0.8}Sc_{0.2}N$ thin films on Si wafer; (b) top Pt electrodes patterning by lift-off; (c) $SiO_2$ hard mask deposition and patterning; (d) $Al_{0.8}Sc_{0.2}N$ film patterning; (e) Si wafer backside thinning; (f) Si backside etch through by $XeF_2$; (g) Pt bottom electrode deposition from the backside; (h) $SiO_2$ hard mask removal.

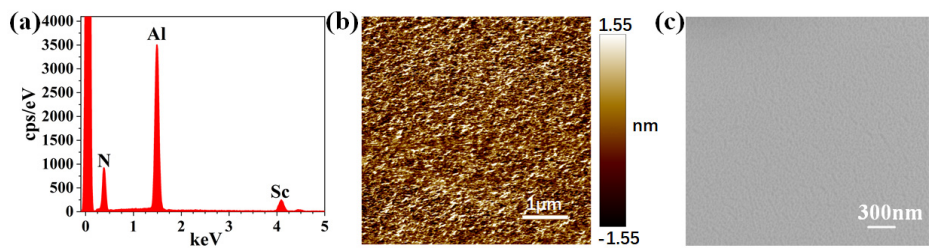

**Fig. 9.** (a) The EDS spectrum, (b) the AFM plot, and (c) surface SEM image of the $Al_{0.8}Sc_{0.2}N$ thin film.

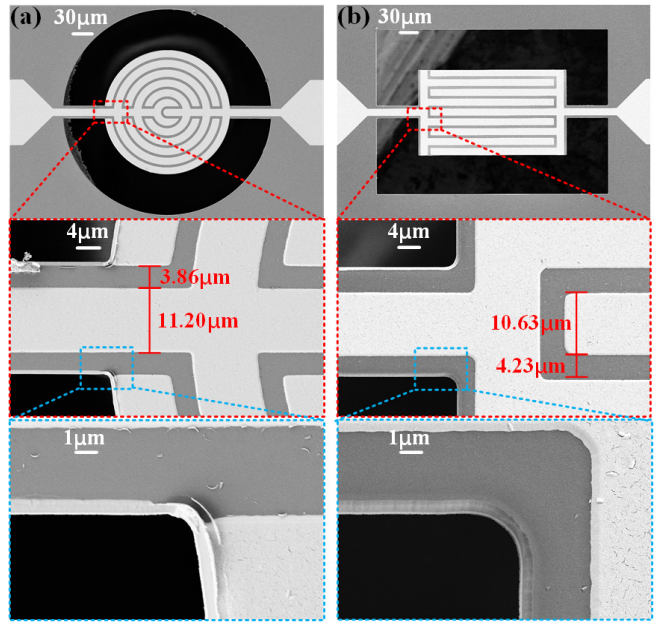

**Fig. 10.** The SEM images of the fabricated (a) circular crested and (b) straight crested LWRs.

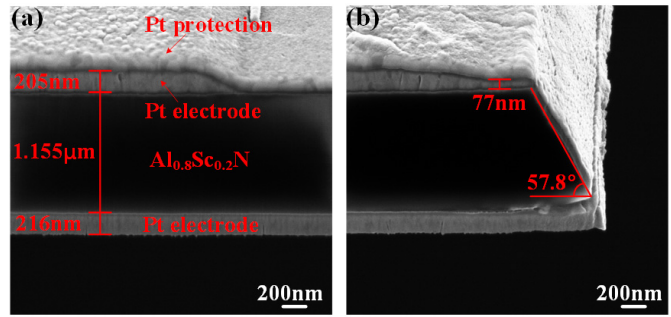

**Fig. 11.** The HIM cross-sectional images of the fabricated LWR.

## 5. Device Performance Characterization

The admittances of the fabricated straight crested and circular crested LWRs are measured by a vector network analyzer (VNA) (Rohde & Schwarz ZNA43) and a probe station (Cascade MSP150) at atmospheric pressure and room temperature and fitted to the mBVD model, which is shown in Figure 12. The resonant frequencies of the straight crested and circular crested LWRs are 215.32 MHz and 216.88 MHz respectively, which are in agreement with the simulation results.

The experimental spurious modes of the circular crested LWR and straight crested LWR are compared to assess the advantage of the presented circular crested LWR. As shown in Figure 12(a), the straight crested LWR suffers from inherent spurious modes that have been analyzed by FEA simulation. Two main spurious modes are observed in the experiment for straight crested LWR at 212.20 MHz and 218.18 MHz, which correspond to the simulated transverse spurious modes at 205.5 MHz and 212.0 MHz respectively. For convenience, the two spurious modes are denoted as A and B respectively. Notably, compared with Figure 7(a), in Figure 12(a), the frequency difference between spurious mode A and $f_s$ is slightly larger and the frequency difference between spurious mode B and $f_s$ is slightly smaller. Moreover, the two transverse spurious modes have higher peaks in the measurement results than in the simulation results. Besides, the small simulated transverse spurious mode at 219.8 MHz is suppressed in the measurement results. A possible reason is that tilted sidewalls shown in Figure 11 modify the boundary condition in the transverse direction, affecting the transverse wavenumbers and damping of transverse spurious modes, thus influencing their frequencies and $Q$ and even suppressing small spurious modes [49]. The boundary condition modification caused by fabrication imperfections may also result in the unpredicted spurious mode of the circular crested LWR shown in Figure 12(b), which is not demonstrated in the simulation result shown in Figure 7(b). On the other hand, as shown in Figure 12(b), compared with the straight crested LWR, the fabricated circular crested LWR successfully suppresses the transverse spurious mode closest to $f_p$, which is in agreement with the simulation result shown in Figure 7(b) and thus verifies the transverse spurious mode suppression capability of the circular crested LWR.

Furthermore, the experimental $Q_{fit}$ of the circular and straight crested LWRs is compared to assess the $Q$ enhancement ability of the presented circular crested LWR. The fabricated circular crested LWR shows a $Q_{fit}$ of 299.9, which is 40.7% higher than that of the fabricated straight crested LWR. The experimental results demonstrate the effectiveness of circular crested LWRs in $Q$ enhancement compared with straight crested LWRs. However, compared with the simulated results, the measured $Q_{fit}$ is degraded, as unwanted geometry deviations emerge during the fabrication process. As shown in Figure 10, rounded instead of right-angle corners are at the junctions of tethers and the resonance cavity due to the nonideal lithography and RIE step. Moreover, tilted sidewalls are observed in Figure 11(b), which are attributed to the nonideal ICP-RIE process. The rounded corners and tilted sidewalls can significantly reduce $Q$ [25, 42]. Besides, the electrical connections required for measurement can lead to additional electrical energy loss. As for the $k_{eff}^2$, the straight crested and circular crested LWR show comparable $k_{eff}^2$ of 1.405% and 1.617% respectively, which are basically consistent with the simulation results. Although we have tried to include experimental damping sources in the actual-design simulations, discrepancies between the simulation and experimental results are inevitable due to the complexity of experimental imperfections. Thus, continuous iterations between simulations and experiments are required to develop more precise simulation methods in future works. With optimization of the simulation methods, fabrication process, and device designs, the performance of the circular crested LWR can be further improved.

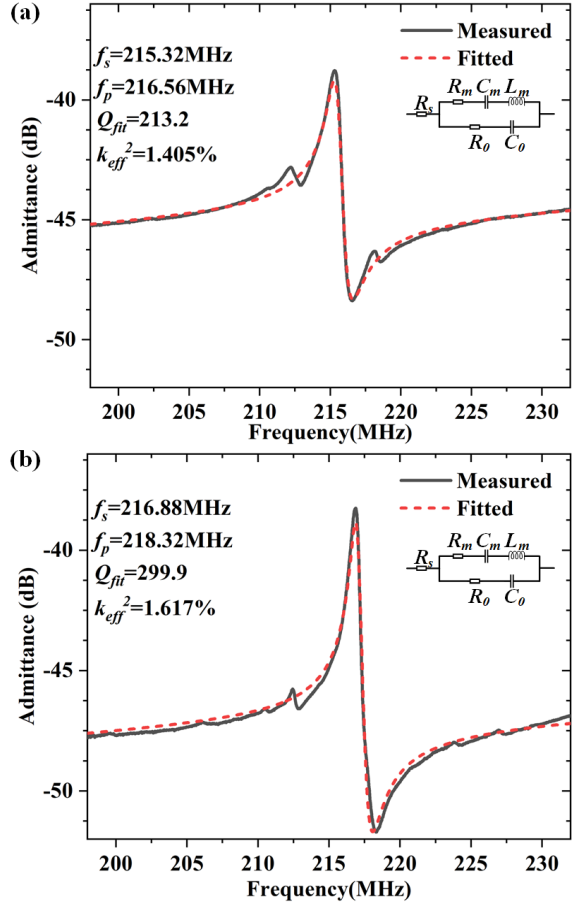

**Fig. 12.** The measured and mBVD fitted admittance of the fabricated (a) straight crested and (b) circular crested LWRs.

## 6. Discussion

In Section 3, it is shown that the simulated performances of the ideal-design and actual-design LWRs are different due to the variation in the resonator structure. The tether width is one of the most important influence factors, as it significantly affects the efficiency of energy dissipation to surroundings and the boundary conditions in the transverse direction. The influence of tether width on conventional straight crested LWRs has been investigated by many researchers [42, 43], yet the influence on circular crested LWR has not been studied. Thus, 3D FEA simulations are employed to investigate the impact of tether width on actual-design circular crested LWRs and the difference from straight crested LWRs, in which tether widths are swept from 1 μm to 35 μm with a wavelength of 30 μm a tether length to wavelength ratio of 2.08. As shown in Figure 13(a), the straight crested LWR suffers from inherent transverse spurious modes adjacent to the main resonance regardless of tether width. With the increase in tether width, transverse spurious modes are first mitigated with tether widths below 35 μm and then slightly enhanced with a tether width of 35 μm, which can possibly be explained by the effect of tethers on the energy dissipation and excitation of the spurious modes. With the increase of tether width, the energy dissipation of transverse spurious modes should increase due to the enhanced anchor loss, which leads to the initial mitigation of the spurious modes [32, 44]. However, wider tethers give rise to undesirable modes, which may explain the enhanced spurious mode with tethers at 35 μm [48]. On the contrary, as shown in Figure 13(b), the circular crested LWR maintains its clean spectrum near the main resonance with tether width ranging from 1 μm to 35 μm, demonstrating that its resistance against transverse spurious modes is barely affected by tether width, which is more advantageous than the straight crested LWRs. Notably, the simulated admittance curves of the actual-design LWRs with 1 μm wide tethers are different from that of the ideal-design LWRs with 1 μm wide tethers shown in Figure 5(a) and (b). Besides the influence of additional damping sources included in the simulation of actual-design LWRs, busbars also significantly influence LWR performances and cause the difference.

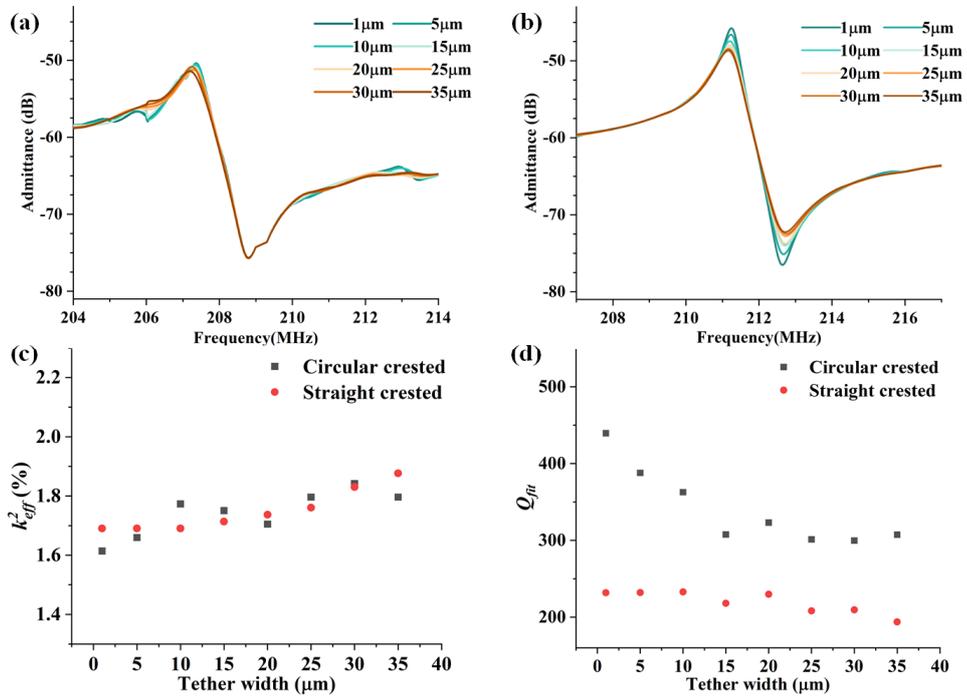

**Fig. 13.** The simulated admittance of the actual-design (a) straight crested and (b) circular crested LWRs with different tether widths; the simulated (c) $k_{eff}^2$ and (d) $Q_{fit}$ of the straight crested and circular crested LWRs with different tether widths.

Figure 13(c) shows the variation of $k_{eff}^2$ of the straight crested and circular crested LWRs with tether width. With the increase in tether width, the simulated $k_{eff}^2$ of the straight crested and circular crested LWRs remains approximately unchanged, showing that the tether width merely affects the effective coupling efficiency of the S0 mode. Figure 13(d) presents the relationship between $Q_{fit}$ and tether width. With the increase of tether width, $Q_{fit}$ of the straight crested LWR shows a general trend of slow decline. Compared with the straight crested LWR, the circular crested LWR shows a more rapid decrease in $Q_{fit}$ as the tether width increases from 1 μm to 10 μm, indicating that the anchor loss of the circular crested LWR may be more sensitive to the tether width when tethers are narrow compared with wavelengths. Notably, with tether width ranging from 1 μm to 35 μm, the circular crested LWR exhibits significantly higher simulated $Q_{fit}$ than the straight crested LWR, illustrating that the effectiveness of employing circular crested Lamb waves in reducing anchor loss can be adapted to resonators with various tether designs.

## 7. Conclusion

For the first time, we report a novel design of circular crested LWR. The S0 mode, which is the most representative Lamb wave mode, is comprehensively studied as an instance to present the advantages of the circular crested LWR. It is demonstrated that the circular crested LWR utilizes circular crested Lamb waves to avoid the existence of the transverse direction and thus diminishes transverse spurious modes, while conventional straight crested LWRs inherently suffer from transverse spurious modes due to the insufficient wave guiding in the lateral direction. Moreover, compared with straight crested Lamb waves employed by conventional straight crested LWRs exhibiting the same displacement amplitudes along the propagation direction, circular crested Lamb waves employed by the proposed circular crested LWR exhibit gradually dwindling displacement toward device edges, which can effectively concentrate energy in the device center, reduce anchor loss, and improve $Q_{fit}$. Besides, the design of overlapping the lateral outer edges of the resonance cavity and the outermost IDT fingers enhances the robustness against misalignment in the fabrication process. Fabricated with an $Al_{0.8}Sc_{0.2}N$ piezoelectric thin film, the circular crested LWR achieves effective transverse spurious mode suppression and a $Q_{fit}$ enhancement of 40.7% compared with the straight crested LWR with no $k_{eff}^2$ degradation when operating in S0 mode. The novel design of the circular crested LWR reported in this work may benefit the development of high-performance filters and oscillators with small footprints.

### Author contributions

### Competing interests

The authors declare that they have no known competing financial interests.